\documentclass[article]{aastex631}

\newcommand{\HI}{\ion{H}{1}~}
\newcommand{\Halpha}{H$\alpha$}
\newcommand{\Lya}{Ly$\alpha$}
\newcommand{\kms}{km s$^{-1}$}
\usepackage{gensymb}
\usepackage{lipsum}
\usepackage[flushleft]{threeparttable}

\makeatletter
\DeclareRobustCommand{\HI}{%
  \mbox{H\check@mathfonts\fontsize\sf@size\z@\selectfont I}%
}
\makeatother

\shorttitle{Anomalous HI Clouds in M\,100}

\begin{document}

\title{DIISC-I: The Discovery of Kinematically Anomalous HI Clouds in M\,100}

\correspondingauthor{Hansung B. Gim}
\email{hansung.gim@montana.edu}

\author[0000-0003-1436-7658]{Hansung B. Gim}
\affiliation{School of Earth \& Space Exploration, Arizona State University, 781 E Terrace Mall, Tempe, AZ 85287, USA}
\affiliation{Cosmology Initiatives, Arizona State University, 650 E Tyler Mall, Tempe, AZ 85287, USA}
\affiliation{Department of Physics, Montana State University, P.O. Box 173840, Bozeman, MT 59717, USA}

\author[0000-0002-2724-8298]{Sanchayeeta Borthakur}
\affiliation{School of Earth \& Space Exploration, Arizona State University, 781 E Terrace Mall, Tempe, AZ 85287, USA}

\author[0000-0003-3168-5922]{Emmanuel Momjian}
\affiliation{National Radio Astronomy Observatory, 1003 Lopezville Rd, Socorro, NM 87801, USA}

\author[0000-0002-3472-0490]{Mansi Padave}
\affiliation{School of Earth \& Space Exploration, Arizona State University, 781 E Terrace Mall, Tempe, AZ 85287, USA}

\author[0000-0003-1268-5230]{Rolf A. Jansen}
\affiliation{School of Earth \& Space Exploration, Arizona State University, 781 E Terrace Mall, Tempe, AZ 85287, USA}

\author[0000-0001-8421-5890]{Dylan Nelson}
\affiliation{Universität Heidelberg, Zentrum f\"ur Astronomie, Institut f\"ur theoretische Astrophysik, Albert-Ueberle-Str. 2, 69120 Heidelberg, Germany}

\author[0000-0001-6670-6370]{Timothy M. Heckman}
\affiliation{Center for Astrophysical Sciences, Department of Physics and Astronomy, Johns Hopkins University, Baltimore, MD 21218, USA}

\author[0000-0001-5448-1821]{Robert C. Kennicutt, Jr.}
\affiliation{Department of Astronomy and Steward Observatory, University of Arizona, Tucson, AZ 85721, USA}
\affiliation{Department of Physics and Astronomy, Texas A\&M University, College Station, TX 77843, USA}
\affiliation{Institute of Astronomy, University of Cambridge, Madingley Road, Cambridge CB3 0HA, UK}

\author[0000-0003-0724-4115]{Andrew J. Fox}
\affiliation{AURA for ESA, Space Telescope Science Institute, 3700 San Martin Drive, Baltimore, MD 21218, USA}

\author[0000-0001-8898-2800]{Jorge L. Pineda}
\affiliation{Jet Propulsion Laboratory, California Institute of Technology, 4800 Oak Grove Drive, Pasadena, CA 91109-8099, USA}

\author[0000-0002-8528-7340]{David Thilker}
\affiliation{Department of Physics \& Astronomy, Johns Hopkins University, Baltimore, MD, 21218, USA}

\author{Guinevere Kauffmann}
\affiliation{Max-Planck Institut f\"ur Astrophysik, D-85741 Garching, Germany}

\author[0000-0002-7982-412X]{Jason Tumlinson}
\affiliation{Center for Astrophysical Sciences, Department of Physics and Astronomy, Johns Hopkins University, Baltimore, MD 21218, USA}
\affiliation{Space Telescope Science Institute, 3700 San Martin Drive, Baltimore, MD 21218, USA}

\begin{abstract}

We report the discovery of two kinematically anomalous atomic hydrogen (\HI) clouds in M\,100 (NGC\, 4321), which was observed as part of the Deciphering the Interplay between the Interstellar medium, Stars, and the Circumgalactic medium (\mbox{DIISC}) survey in \HI\ 21\,cm at 3.3 \kms\ spectroscopic and 44\arcsec$\times$30\arcsec spatial resolution using the Karl G. Jansky Very Large Array \footnote{The National Radio Astronomy Observatory is a facility of the National Science Foundation operated under cooperative agreement by Associated Universities, Inc.}. 
These clouds were identified as structures that show significant kinematic offsets from the rotating disk of M100. 
The velocity offsets of 40 \kms\ observed in these clouds are comparable to the offsets seen in intermediate-velocity clouds (IVCs) in the circumgalactic medium (CGM) of the Milky Way and nearby galaxies. 
We find that one anomalous cloud in M\,100 is associated with star-forming regions detected in H$\alpha$ and far-ultraviolet imaging. Our investigation shows that anomalous clouds in M\,100 may originate from multiple mechanisms, such as star formation feedback-driven outflows, ram-pressure stripping, and tidal interactions with satellite galaxies. Moreover, we do not detect any cool CGM at 38.8~kpc from the center of M\,100, giving an upper limit of N(\HI) $\le$ 1.7$\times$10$^{13}$~cm$^{-2}$ (3$\sigma$). Since M\,100 is in the Virgo cluster, the non-existence of neutral/cool CGM is a likely pathway for turning it into a red galaxy.

\end{abstract}


\section{Introduction} \label{sec:intro}

The buildup of stellar mass in a galaxy is a consequence of the balance among gas supply, removal, and consumption; therefore it is crucial to trace gas flows in and out of galaxies to understand their growth \citep{sancisi08, putman12}. Deep observations of the 21~cm line emission from atomic hydrogen (\HI) provide a useful tracer of gas flows for exploring the accretion or outflow of atomic gas in and out of galaxies. In general, gas flows (accretion or outflows) are expected to exhibit kinematics that are inconsistent with the galaxy's rotation, thus showing up as spurs or bridges in the position-velocity (PV) diagram \citep[][and references therein]{sancisi08}.

Discrete clouds of neutral hydrogen gas with peculiar kinematics relative to the organized rotation of a galaxy disk, which we will call anomalous \HI\ clouds (AHCs) hereafter, represent gas that may be falling into  or is being ejected out of the disk. In terms of the velocity offset, AHCs are analogous to the intermediate- and high-velocity clouds (IVCs \& HVCs) in the Milky Way Galaxy \citep{wakker97}. Since the first discovery of an HVC in the 21~cm emission line \citep{muller63}, several studies have established that HVCs have characteristic velocities of $-300 \lesssim v_{GSR} \lesssim 300$~km s$^{-1}$, typical velocity widths of 20--30~km s$^{-1}$, \HI\ masses of 0.1--5$\times 10^{6}$ M$_{\odot}$, and physical sizes of $\sim$15~kpc \citep[see the review by ][]{putman12}.

AHCs have been identified in several nearby galaxies. Even though they were referred by different names across individual studies, they do all fall under the same definition of AHC; henceforce will be referred to as AHC. AHCs were detected in face-on galaxies such as M~101 \citep{vanderhulst88}, NGC~628 \citep{kamphuis92}, and NGC~6946 \citep{kamphuis93, boomsma08} by their large velocity offsets from the rotating \HI\ disk of $| \delta v |$ $\approx$130--150, $\approx$100, and $> $50 \kms\, respectively. Each of these studies led to a different conclusion about the origins of the AHCs, e.g., collisions with tidally stripped gas clouds for M~101, disk perturbation due to gas accretion in NGC~628, and star formation feedback in NGC~6946. Similarly, eight AHCs were detected by \citet{miller09} in the thick disk of M~83, where three are in the spiral arms within the optical disk and the remaining are outside the optical disk. They argued that AHCs within the optical disk are associated with star formation-driven feedback while the others result from tidal interactions. AHCs were also observed in inclined ($30 \lesssim i \lesssim 75$\degree) galaxies such as NGC~2403 \citep{fraternali01}, M~31 \citep{thilker04, westmeier05}, and NGC~1003 \citep{heald15}. \citet{fraternali01} identified several clumpy clouds in the slowly rotating halo of NGC~2403, where the slow rotation of the halo appeared as a `beard' in a PV diagram. Twenty discrete \HI~clouds were observed in M~31 \citep{thilker04, westmeier05} with \HI\ masses of $\rm 10^{5-7} \; M_{\odot}$ and velocity offsets of $90 \lesssim |\delta v| \lesssim 212$~\kms. These clouds are likely tracing tidally stripped gas from recent or ongoing mergers \citep{thilker04}. In addition, the edge-on galaxy NGC~891 has anomalous velocity clouds identified as counter-rotating gas clouds with velocity offsets of $\delta v \sim$100~\kms \citep{oosterloo07}.

Specific simulations to understand the nature and origin of the HVCs in the Milky Way \citep{sommer-larsen06, peek08} have reproduced the observed HVCs as gas condensing out of the hot halo at large distances of $\sim$100~kpc from the galactic center, which then rains down toward the disk. Recent studies by \citet{fraternali15} and \citet{marasco17} find that the Milky Way's complex C and Smith clouds are consistent with outflowing material that has condensed and is returning back to the disk. Whatever their origins, HVCs are expected to be short-lived with typical lifetimes of a few 100~Myr  \citep{heitsch09, armillotta17} and hence, the galaxies would need to generate them constantly to match the observed numbers. Several studies using the large-scale cosmological hydrodynamic simulations investigated the gas properties in galaxies. Studies with the IllustrisTNG100 explored the stripping of the jellyfish galaxies in massive groups/clusters \citep{yun19}, the asymmetries of \HI\ distributions in galaxies \citep{watts20}, and the environmental effects on the \HI\ properties \citep{stevens19}. \citet{bahe16} and \citet{crain17} exploited the EAGLE simulation \citep{schaye15} to investigate the global properties of \HI\ gas such as radial distributions, \HI\ mass function, morphologies, and size. However, those studies did not trace individual gas clouds due to the limited resolution. More recently, IllustrisTNG50 showed the possibilities to explore gas clouds with better mass resolution of 8$\times$10$^{4}$~M$_{\odot}$ and spatial resolution of $\sim$100--140~pc \citep{nelson19}. This study explored the multi-phase outflows including those with cool gas components, while \citet{nelson20} analyzed the numerous sub-kpc sized individual clouds and found to populate the hot CGM of massive halos in IllustrisTNG50. These studies implied that the properties of AHCs in galaxies will be studied in the context of large-scale cosmological hydrodynamic simulations.

To probe the flow of gas into and out of the disk of galaxies in a holistic fashion, we have designed the Deciphering the Interplay between the Interstellar medium, Stars, and the Circumgalactic medium (\mbox{DIISC}) survey aimed to trace the entire pathway of the baryons from the circumgalactic medium (CGM) through the interstellar medium (ISM) to young and old stars. The \mbox{DIISC} sample consists of 35 low-redshift ($z= 0.002-0.05$) disk galaxies that have an ultraviolet (UV) bright Quasi-Stellar Object (QSO) within 3.5 times their \HI\ radius \footnote{The \HI\ radius is defined as the mean radius deprojected at a mass surface density $\rm\Sigma(M_{HI}) = 1~M_{\odot}~pc^{-2}$}. This enables us to probe the CGM in absorption with the Cosmic Origins Spectrograph \citep[COS, ][]{green12} aboard the \emph{Hubble Space Telescope} (HST). We combine that with deep \HI~21cm imaging to trace the ISM with the {\it NSF's Karl G. Jansky Very Large Array} (VLA). We identify regions with young ($\lesssim$100\,Myr) stellar populations in archival far-ultraviolet (FUV) images from the \emph{Galaxy Evolution Explorer} (GALEX), and regions with current massive star formation ($\lesssim$10\,Myr) in new H$\alpha$ images acquired at the \emph{Vatican Advanced Technology Telescope} (VATT). In this paper, we present the first results from the VLA \HI\ imaging obtained as part of the \mbox{DIISC} program, i.e., the discovery of kinematically anomalous \HI\ clouds in M\,100. The design of the \mbox{DIISC} survey will be described in detail in Borthakur et al (in prep.).

M\,100, also known as NGC~4321, is a grand design spiral galaxy in the Virgo cluster with a stellar mass of $\rm 4.83 \times 10^{10}$~M$_{\odot}$ and a star formation rate (SFR) of 3.23 M$_{\odot}$ yr$^{-1}$ \citep{hunt19} at a projected distance of 783~kpc from the center of the cluster. The galaxy is close to face-on, with an inclination with respect to our line-of-sight $i \approx$30\degree. Since the first \HI\ interferometric observations with the VLA \citep{vangorkom85} and the Westerbork Synthesis Radio Telescope \citep{warmels88}, several studies revisited M\,100 with different spatial and spectral resolutions \citep{cayatte90, knapen93, haan08, chung09}. Those studies reported an \HI\ mass of 2.6--3.4$\times 10^{9}$~M$_{\odot}$, velocity width of $W_{50}=$238--244~\kms, and a systemic velocity of 1570--1580~\kms. In an extensive study of M\,100 with the VLA, \citet{knapen93} showed that its \HI\ distribution is asymmetric. They discovered a large-scale deviation in the velocity field along the South-West (SW) extension showing an offset of the observed velocity field with respect to that expected from the undisturbed disk model. They discussed that this velocity deviation might be a consequence of a perturbation of the disk due to the close passage of the dwarf companion, NGC~4322. Wiggles observed in the velocity field at the SW extension were interpreted as a gas streaming motion due to the spiral arm's density wave. At the velocity resolution of their data (20.8 \kms), \citeauthor{knapen93} did not detect any AHCs in M\,100.

In this paper, we present the discovery of two AHCs in  M\,100 based on our new observations at an unprecedented spectral resolution of 3.3~\kms. We describe our data acquisition and reduction of VLA \HI\ observations, HST-COS QSO absorption spectroscopy, and GALEX FUV and VATT H$\alpha$ imaging in \S\ref{sec:obs}. We present our analysis of the \HI\ properties of M\,100 and the discovery of two AHCs in \S\ref{sec:hi_m100}, followed by a discussion of the nature and  origin of these AHCs in \S\ref{sec:discussion}. We conclude with a brief summary in \S\ref{sec:conclusion}. In this work, we adopt the luminosity distance for M\,100 of 13.93$\pm$0.12 Mpc, as provided by the Cosmicflows-3 database \citep{cosmic_flow3}, giving a linear scale of ($14\farcs80\pm0.13$)\,kpc$^{-1}$ \footnote{We note that a recent estimate of the distance to M100 is 15.44$\pm$1.62~Mpc \citep{anand21}. While the conclusions of the paper are not affected, the values of the derived parameters can have a change of $\sim$20\%.}

\section{Observations and Data} \label{sec:obs}
\subsection{VLA observations} \label{sec:hiobs}
We observed the \HI\ 21\,cm hyperfine transition in M\,100 with the VLA in D-configuration as part of the \mbox{VLA-DIISC} program (Project code of 18A-006, P.I: S.~Borthakur). Between 2018 September and 2018 November, we accrued a total integration time of 4~hours, with a correlator set up that delivered 2048 channels for a total bandwidth of 16~MHz resulting in a channel width of 7.812~kHz, equivalent to the velocity spacing of 1.65 \kms\ at the M\,100's distance.

The data reduction was performed in CASA 5.6.1 \citep{mcmullin07} following standard VLA data reduction, editing, and calibration steps. The source 3C286 was used to calibrate the delay, bandpass, and the absolute flux density scale. J1254+1141 served as the phase calibrator for amplitude and phase gain calibrations. Hanning smoothing was applied to reduce Gibbs ringing. We subtracted the continuum from the calibrated data in the uv-plane using the CASA task \textit{uvcontsub}. Then, those data were imaged using \textit{tclean} with a cell size of 4\arcsec$\times$4\arcsec~and an image size of 800$\times$ 800 pixels. Primary beam correction was applied. The cleaning was performed down to 1.5 times the local root-mean-squared (RMS) noise. Since the beam size varies as a function of frequency within the 16 MHz bandpass, we smoothed the image cube to the same synthesized beam size across all channels using task \textit{imsmooth}. We made two image cubes with different weightings on the baselines: uniform weighting to achieve the best angular resolution and natural weighting to enhance low surface brightness features. 
Our data cubes achieve RMS noise of 1.7~mJy per channel over the beam size of 44\arcsec$\times$30\arcsec~in the uniformly-weighted cube and 1.3 mJy per channel over the beam size of 73\arcsec$\times$52\arcsec~in the naturally-weighted cube. 
Both \HI\ cubes have a velocity resolution of 3.3~km s$^{-1}$ as a consequence of Hanning smoothing.

We used the Source Finding Application-2 \citep[SoFiA-2, ][]{serra15} to extract the \HI\ 21~cm emission of M\,100 above 3$\sigma$ significance, where $\sigma$ is the RMS noise of $1.7$~mJy beam$^{-1}$ per channel and $1.3$~mJy beam$^{-1}$ per channel for the uniformly- and naturally-weighted cubes, respectively. We identified sources using the Smooth$+$Clip algorithm \citep{serra12} with spatial averaging using 1$\times$1, 3$\times$3, and 5$\times$5 pixel Gaussian kernels, and with spectral averaging using 1, 3, and 5 channels boxcar kernels. Because the spatial averaging is performed within the synthesized beamsize, this process does not distort the data cube. We generated moment maps within velocities of 1412--1717 \kms\ (corresponding to the channel numbers 41--226) for two weighting schemes. Corresponding \HI\ column density maps were produced by dividing the velocity-integrated flux density (total intensity or the 0th moment), $I_{\rm tot}$ in $\rm Jy \; beam^{-1} \; km \; s^{-1}$ by major and minor axis sizes of the beam ($\theta_{\rm maj}$ and $\theta_{\rm min}$ in arcsec),

\begin{equation}
N_{HI} = 1.1 \times 10^{24} {{I_{tot}} \over {\theta_{maj} \theta_{min}}} 
\end{equation}
where the constant of proportionality ensures units of cm$^{-2}$ for the column density, $N_{\rm\HI}$. \HI\ masses ($M_{\rm\HI}$ in M$_{\odot}$) were calculated by integrating the flux density ($S_{\nu}$ in Jy) over velocity $dv$, multiplied by the luminosity distance, $d_{\rm L}$ of M\,100, and with the appropriate unit conversion assuming that all of the \HI\ is optically thin, 

\begin{equation}
M_{HI} = 2.35 \times 10^{5} d_{L}^{2} \int S_{\nu} dv
\label{eq:himass}
\end{equation}

\subsection{QSO absorption spectroscopy} \label{sec:cos_obs}

The inner CGM of M\,100 was probed at an impact parameter\footnote{The projected separation of the QSO in the rest-frame of M100} of 38.8~kpc via UV absorption spectroscopy of the background QSO SDSS\, {J122330.78}\-{+154507.3} using the Cosmic Origins Spectrograph (COS) aboard the \emph{Hubble Space Telescope}. Observations with the G130M grating and a total exposure time of 10440~sec were carried out on UT 2016 March 26 as part of the \mbox{COS-DIISC} Survey \footnote{The \mbox{COS-DIISC} survey is a large HST program that aims to trace the disk-CGM interface in 35 low-redshift disk galaxies using UV bright background QSO (Borthakur et al. in prep.).} (GO 14071; PI: S. Borthakur).  
The spectra cover the  wavelength range of 1142--1443\AA\ in the observed frame corresponding to 1136--1435\AA\ in the rest frame of M\,100 at a  resolution of $R$\,$\sim$\,15000 ($\sim$20 \kms). Our data cover line transitions such as \ion{H}{1} $\lambda$1216 (\Lya), \ion{Si}{2} $\lambda \lambda$ 1190, 1193, 1260, \ion{Si}{3} $\lambda$1206,  \ion{Si}{4} $\lambda\lambda$1393, 1402, \ion{C}{2} $\lambda$1334, and \ion{N}{5} $\lambda\lambda$1239, 1242. The data were calibrated and reduced using the standard COS pipeline \citep[described in the COS Data Handbook; ][]{rafelski18}. The extracted spectra were normalized by a continuum that was visually identified using absorption-free regions within $\pm 1500$~\kms~from each absorption line center. The continuum was then estimated by fitting a Legendre polynomial of order between 1 and 5, similar to the procedure used by \cite{sembach04}. This corrected for any low-order variations of the QSO flux near the position of the lines. The spectrum was then divided by the continuum to get the normalized spectrum.

\begin{figure}[!htb]
    \centering
    \includegraphics[trim = 0cm 0mm 0cm 0cm, clip,scale=0.5]{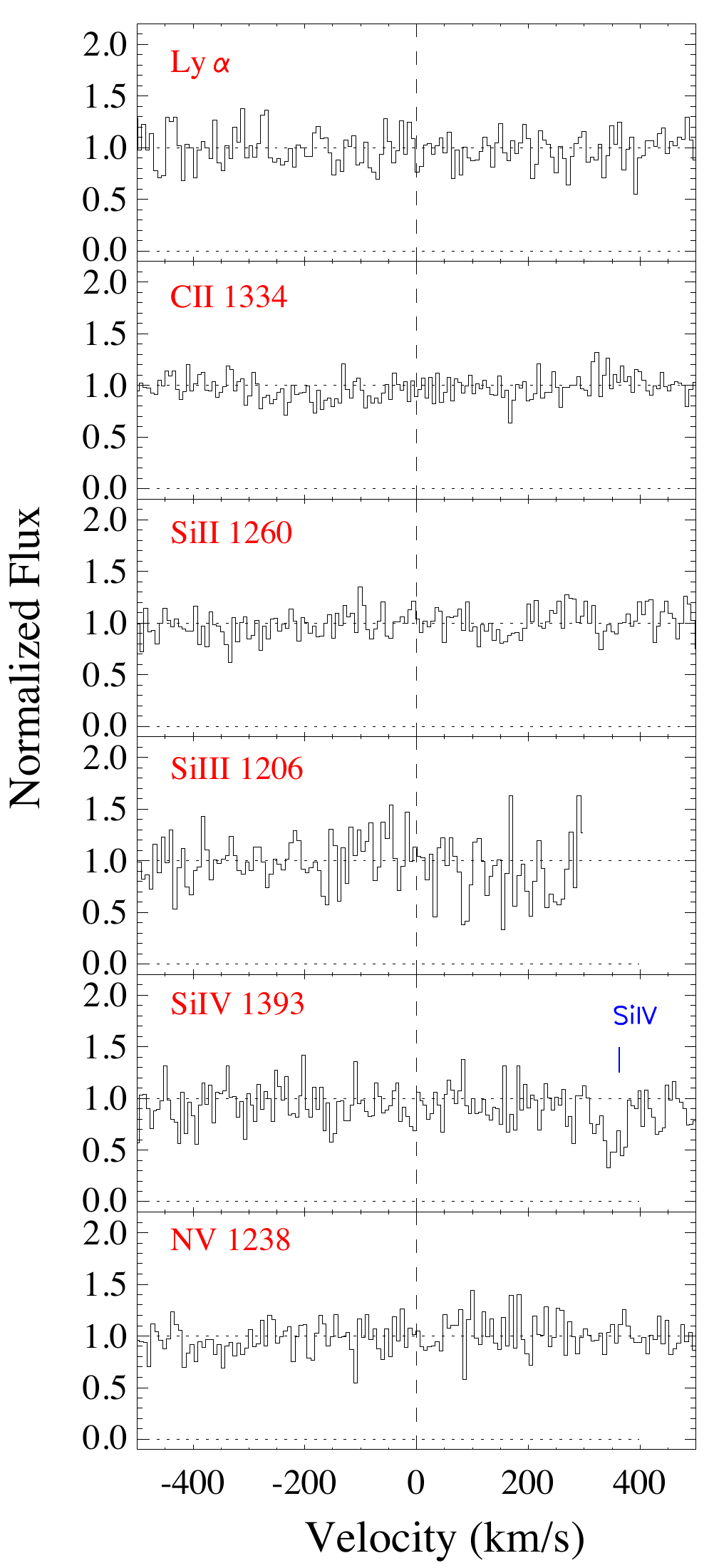}
    \caption{ COS FUV spectra of the QSO~J122330$+$154507 probing the circumgalactic medium of M~100 at a galactocentric distance of 38.8~kpc. We do not detect any species within the rest-frame wavelength coverage of 1136–1435~$\rm \AA$. We present the limiting column densities for \ion{H}{1} $\lambda$ 1215, \ion{Si}{2} $\lambda\lambda$ 1260,  \ion{C}{2} $\lambda$1334, \ion{Si}{3} $\lambda$1206, \ion{Si}{4} 
    $\lambda\lambda$1393, and \ion{N}{5} 1238 in Table~\ref{abs_lines}. Intervening Milky Way absorption due to \ion{Si}{4} 1402 is shown in blue. }
    \label{fig:cos_spec}
\end{figure}

We did not detect any UV absorption lines associated with the CGM of M\,100 in the spectra of QSO\,{J122330}\-{+154507} at an impact parameter of 38.8~kpc corresponding to 0.14~$R_{vir}$, where $R_{vir}$ is the virial radius of 280~kpc corresponding to stellar mass, log~$M_{\star}=10.68$ \citep[estimated using the methodology described in ][]{borthakur15} (Figure~\ref{fig:cos_spec}). We estimated 3$\sigma$ upper limits for the equivalent width and column density for the spectral region within $\pm$50 \kms\ of the center of each transition (see Table~\ref{abs_lines}). The atomic data is adapted from \citet{morton03}. The non-detection of \Lya, \ion{N}{5} or any other species with ionization potential in-between them indicates that the circumgalactic gas around this galaxy is at a temperature higher than 2$\times$10$^{5}$~K, where \ion{N}{5} is expected to peak for a collisionally ionized system.

\begin{deluxetable*}{ccccc}   
\label{abs_lines}
\tabletypesize{\scriptsize}
\tablecaption{Upper limits to absorption lines probing the CGM of M\,100 in the spectra of QSO~J122330$+$154507}
\tablewidth{6pt}
\tablehead{
  \colhead{Species} & \colhead{$\lambda_{rest}$}  & \colhead{$\lambda_{obs}^{a}$} & \colhead{$W_{rest}^{b}$} & \colhead{log N$^{c}$}  \\
          \colhead{} & \colhead{($\rm \AA$)} &  \colhead{($\rm \AA$)} & \colhead{($\rm m\AA$)}  &\colhead{log(cm$^{-2}$)}} 
\startdata
 \Lya$^{b}$          &  1215.67    &       1222.06   &    $\le$ 93.6      &  $\le$ 13.24  \\
 \ion{Si}{2}           &  1260.42    &       1267.04   &    $\le$ 78.3      &  $\le$ 12.52  \\
 \ion{Si}{3}           &  1206.50    &       1212.84   &    $\le$ 222.0    &  $\le$ 13.01  \\
 \ion{Si}{4}           &  1393.76    &       1401.08   &    $\le$ 118.2    &  $\le$ 13.13  \\
 \ion{C}{2}            &  1334.53    &       1341.54   &    $\le$ 48.0      &  $\le$ 13.38  \\
 \ion{N}{5}            &  1238.82    &       1245.33   &    $\le$ 81.0      &  $\le$ 13.58  \\
\enddata
\tablenotetext{a}{The expected observed wavelength of the absorption lines for a redshift of M\,100, z$=0.005254$}
\tablenotetext{b}{Limiting equivalent width represents the 3$\sigma$ noise over $\pm$50~km~s$\rm ^{-1}$ from the expected line center.}
\tablenotetext{c}{Limiting column density is estimated assuming a linear conversion between  equivalent width and column density. }
\end{deluxetable*}

\subsection{Far-Ultraviolet and H$\alpha$ Surface Photometry} \label{sec:nuv_obs}

We retrieved pipeline-processed archival GALEX \citep{martin05, morrissey05} FUV imagery covering M\,100, taken as part of the Nearby Galaxy Survey \citep[NGS; ][]{bianchi03, gildepaz04, gildepaz07} and All-Sky Imaging Survey from MAST\footnote{https://mast.stsci.edu/}. We stacked the individual images weighted by the exposure time, excluding background flux measured in source-free regions away from the galaxy. The resulting FUV image of M\,100 has an effective exposure time of 6247\,s, covers 15\arcmin\ $\times$15\arcmin\ at a plate scale of 1$\farcs$5 pixel$^{-1}$, and has a resolution of $\sim$6\arcsec (FWHM). The Galactic (foreground) extinction, adopting the \citet{schlafly11} recalibration of the \citet{schlegel98} dust map, and assuming the extinction curve of \citet{cardelli89} with $R_V = 3.1$, was estimated to be $A_{\rm FUV}$ =7.29 $\times$ $E(B-V)$ = (0.166 $\pm$ 0.002) mag. After correction for extinction, we reach a 1$\sigma$ surface brightness sensitivity limit of 4.63$\times$10$^{-19}$ erg\,s$^{-1}$\,cm$^{-2}$\,{\AA}$^{-1}$.

We observed M\,100 with the VATT4k CCD imager of the 1.8\,m Vatican Advanced Technology Telescope (VATT) at the Mt.~Graham International Observatory in SDSS r-band and  narrow-band H$\alpha$ filters. The VATT CCD imager has a field of view of $\sim$12$\farcm$5$\times$12$\farcm$5 and a plate scale of 0$\farcs$375 pixel$^{-1}$ after 2$\times$2 binning on read-out. The observations were executed on UT 2019 March 31, and comprised 2$\times$600\,s in the $r$ filter and 4$\times$1200\,s with an H$\alpha$ narrow-band interference filter centered at 658 nm and a nominal bandwidth of 5 nm. Data reduction was performed in IRAF \citep{tody93} using standard procedures, that included bias subtraction, and flatfielding using both dome screen and twilight sky flat exposures. The sky level was fit to the sigma-clipped average intensities measured in source-free regions and subtracted. Cosmic ray induced signal was removed using the LACOSMIC algorithm \citep{vandokkum01} as implemented in IRAF\footnote{The LACOSMIC task was downloaded from http://www.astro.yale.edu/dokkum/lacosmic/download.html}. All images were astrometrically aligned using field stars. The $r$ filter images were photometrically calibrated onto the SDSS DR7 photometric system through aperture photometry of unsaturated stars within the field of view, for an assumed atmospheric extinction coefficient of 0.08 mag airmass$^{-1}$. The photometric zeropoint of the \Halpha\ images was determined from that of the calibrated $r$ filter images using the ratios of the count rates for the \Halpha\ and $r$ images of field stars. 
We also applied the correction for the contribution from [\ion{N}{2}] to \Halpha\ since the [\ion{N}{2}] emission line falls within the bandwidth of the \Halpha\ filter. Adopting an absolute magnitude $\rm M_{B}=-28.6$ for M\,100 \citep{dale07} and the correlation between the [\ion{N}{2}]/\Halpha\ ratio and $\rm M_{B}$ from \citet{kennicutt08}, we assumed the [\ion{N}{2}]/\Halpha\ $=$0.54. The 1$\sigma$ surface brightness sensitivity limit of the final \Halpha\ image, after correction for a Galactic foreground extinction of  $\rm A_{H \alpha}$ = 2.5 $\times$ $E(B-V)$ =(0.057$\pm$0.001) mag, is 2.59$\times$10$^{-22}$ erg\,s$^{-1}$\,cm$^{-2}$\,{\AA}$^{-1}$.

The SFR surface density is estimated through the conversions from FUV and H$\alpha$ intensities as in \citet{lee09}, 

\begin{equation}
  \Sigma {\rm SFR}_{\rm FUV} \; \; \left( {\rm M_{\odot} \; yr^{-1} \; kpc^{-2} }\; \right)= 0.17 \times I_{\rm FUV}  \; \; \left( {\rm MJy \; sr^{-1} } \; \right)
\end{equation}
\begin{equation}
  \Sigma {\rm SFR}_{\rm H \alpha} \; \; \left( {\rm M_{\odot} \; yr^{-1} \; kpc^{-2}} \; \right)= 4.32 \times I_{\rm H \alpha}  \; \; \left( {\rm MJy \; sr^{-1} } \; \right) 
\end{equation}
The 1$\sigma$ uncertainties of the FUV and H$\alpha$ photometry translate to 1$\sigma$ uncertainties of 1.19\,$\times$\,10$^{-4}$ and 5.78\,$\times$\,10$^{-4}$ M$_{\odot}$\,yr$^{-1}$\,kpc$^{-2}$ for SFR surface densities derived from FUV and H$\alpha$, respectively. \\

\begin{figure*}
    \centering
    \includegraphics[angle=0, scale=0.5]{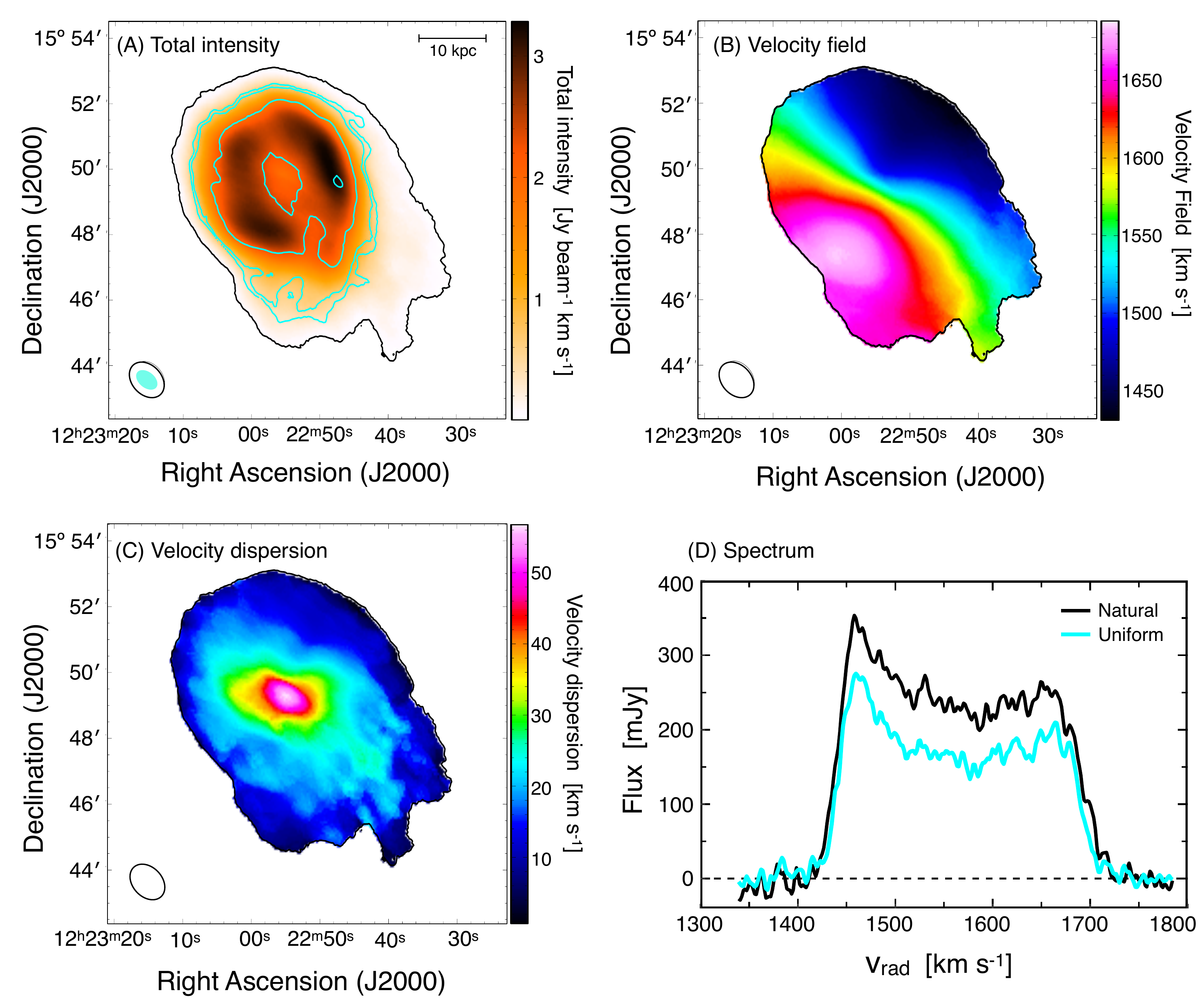}
    \caption{\HI\ 21~cm maps and spectra of M\,100. We present the total intensity map in panel A, the velocity field in panel B, the velocity dispersion in panel C, and the spectra in panel D. These maps are derived from the naturally-weighted cube. We present the total intensity map of the uniformly-weighted cube by cyan contours in panel A. The outermost black and cyan contours indicate the 3$\sigma$ limits of the natural- and uniformly-weighted \HI\ cubes corresponding to $1.85 \times 10^{19}$ and $4.56 \times 10^{19}$~cm$^{-2}$, respectively. The inner cyan contours show the levels corresponding to the column densities of 1, 5, 10$\times 10^{20}$~cm$^{-2}$. The black and cyan ellipses at the bottom left corner in panel A indicate the synthesized beam sizes of the naturally- and uniformly-weighted \HI\ cubes, respectively. The velocity field in panel B shows the overall smooth rotation. The SW extended region has a velocity comparable to the systemic velocity, which is 1565.4 \kms, and low velocity dispersion. The high velocity dispersion at the central region is due to the bar as shown in panel C. The spectra in panel D were extracted from the measurements at naturally (black) and uniformly (cyan)-weighted \HI\ cubes. The \HI\ mass, W$_{50}$, and systemic velocity estimates were derived using the naturally-weighted \HI\ cube.}
    \label{fig:m100_moments}
\end{figure*}

\section{\HI\ properties of M\,100 \label{sec:hi_m100}}

In this section, we describe the \HI\ properties of M\,100. We present the global \HI\ properties and the \HI\ kinematic modeling with the 3D tilted-ring model. Then, we describe our discovery of kinematically distinct AHCs, followed by a description of star formation activity in the vicinity of those structures. 

\subsection{Global \HI\ properties \label{sec:hi}}

\begin{figure*}
    \centering
    \includegraphics[angle=0, scale=0.55]{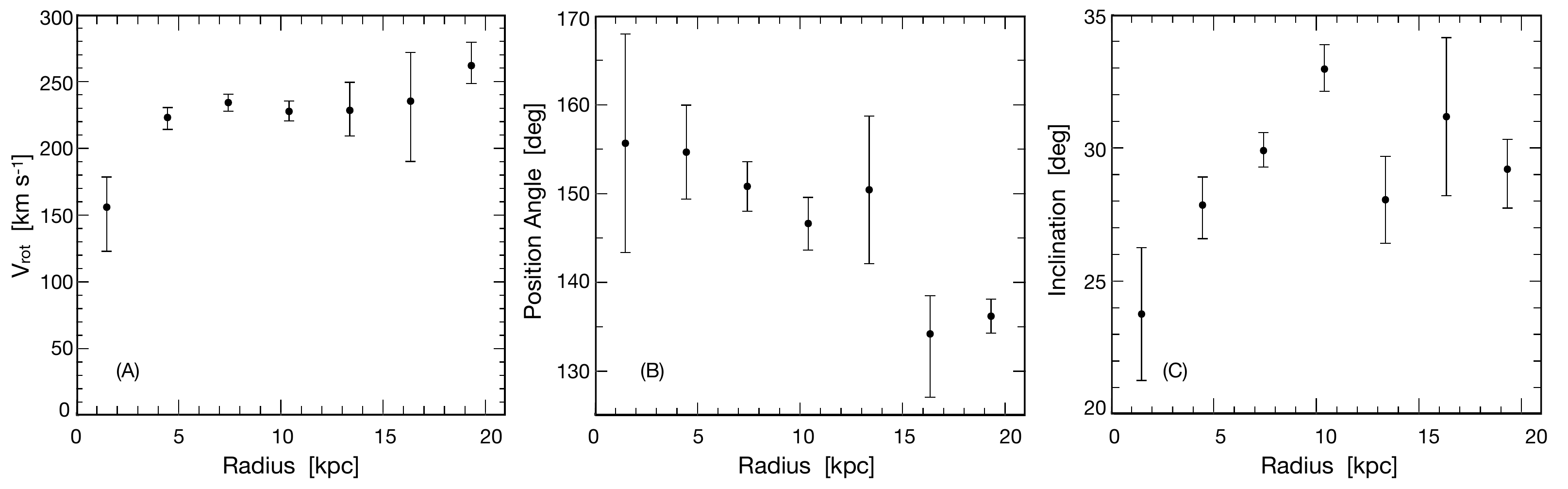}
    \caption{Parameters of the 3D Tilted ring model. We show the rotation velocity in panel (A), position angle in panel (B), and inclination in panel (C) derived from the 3D Tilted-ring model on the image cube. The sudden decrease of the position angle at $\sim 15$~kpc in panel B is due to the lopsidedness of the M\,100 as seen in panel A of Figure~\ref{fig:m100_moments}. }
    \label{fig:m100_3dbarolo}
\end{figure*}

The global \HI\ distributions and kinematics of M\,100 are presented in Figure~\ref{fig:m100_moments} with moment maps and \HI\ spectra. The moment maps in panel A-C were generated by the SoFiA-2 with the naturally-weighted \HI\ cube (the most sensitive one), e.g., the total intensity (moment 0), velocity field (moment 1), and velocity dispersion (moment 2) maps. We show the total intensity maps at two resolutions in panel~A, where the outermost black and cyan contours present the total intensities corresponding to column densities of $1.85 \times 10^{19}$~cm$^{-2}$ in the naturally-weighted map and $4.56 \times 10^{19}$~cm$^{-2}$ in the uniformly-weighted map, respectively. The synthesized beam sizes are presented at the bottom left corner, in black and cyan for the naturally- and uniformly-weighted \HI\ cubes separately. We observe the morphological lopsidedness of the \HI\ distribution, where the extension is toward the SW as seen in the naturally-weighted cube (black line of panel~A). This feature was reported by previous studies \citep{knapen93, chung09}. We measure the major axis of \HI\ disk as 6.9\arcmin~at \HI\ mass surface density, $\rm \Sigma_{HI} = 1 \; M_{\odot} \; pc^{-2}$, which is similar to the major axis of the optical disk of 6.95\arcmin~at B-band \citep{dejong94}. Here, the \HI\ disk extends to a lower column density corresponding to the $\rm \Sigma_{HI} = 0.1~M_{\odot}$~pc$^{-2}$, where its size is 8.4\arcmin. The velocity field map in panel B shows the overall smooth rotation within the radio velocities between 1412$-$1717~km s$^{-1}$ (channels 41-226). The SW extension has a velocity close to the systemic velocity and a low velocity dispersion. We confirm the high velocity dispersion of the gas at the central region of this galaxy previously observed by \citep{knapen93, chung09}, which was attributed to the presence of a bar in M\,100.

In order to account for the entire \HI\ content of M\,100, we used the spectrum of our most sensitive naturally-weighted cube (in black) for the estimates of \HI\ mass, velocity width, and the systemic velocity ($V_{sys}$) in panel D of Figure~\ref{fig:m100_moments}. The total \HI\ mass is estimated to be $\rm M_{HI}= (3.00 \pm 0.02) \times 10^{9} M_{\odot}$ providing the gas mass fraction of 0.058, where gas mass fraction $f_{gas}= M_{HI} / (M_{HI}+M_{*})$ for $\rm M_{*} = 4.83 \times 10^{10} \; M_{\odot}$ \citep{hunt19}. The velocity widths at 50\% ($W_{50}$) and 20\% ($W_{20}$) of the peak flux density are $W_{50}=244.1 \pm 4.7$~km s$^{-1}$ and $W_{20}= 275.2 \pm 4.7$~km s$^{-1}$, respectively. We estimate the systematic velocity of M\,100 as 1565.4$\pm$6.6 km s$^{-1}$ as $V_{sys}= 0.25 \left( V_{20}^{A} + V_{50}^{A}+ V_{50}^{R} + V_{20}^{R} \right)$, where $V_{20}^{A}$ and $V_{20}^{R}$ ($V_{50}^{A}$ and $V_{50}^{R}$) are the approaching and receding velocities of the spectrum at 20\% (50\%) of the peak flux density \citep{chung09}. These values are consistent with previous measurements by \citet{cayatte90}, \citet{gavazzi05}, \citet{haan08}, and \citet{chung09}. We summarize \HI\ properties with stellar mass and SFR in Table~\ref{tab:m100_properties}.

The strong asymmetry of the M\,100's \HI\ disk is visible in the spectra shown in Figure~\ref{fig:m100_moments} (panel D). We quantify the degree of asymmetry using the ratio of area under the \HI\ spectrum divided by the median velocity of 1567.6~km s$^{-1}$ \citep{haynes98}. Its value, $A_{l/h} = 1.2$, indicates strong asymmetry, as only 14\% of local galaxies have $A_{l/h} > 1.2$ \citep{haynes98}.  Analysis of the TNG100 simulation from the IllustrisTNG project concluded that an asymmetric \HI\ distribution is a common phenomenon regardless of environment, even though the degree of asymmetry is weakly dependent on the environment \citep[][]{watts20}. 
However, the observed kinematic lopsidedness does not coincide with the morphological lopsidedness towards the SW extension. 
The asymmetry in the spectrum (panel D) is due to the higher flux at low velocities of $\sim$1450~km s$^{-1}$, implying that more \HI\ is corotating in the NW side of the disk (bluer in the velocity field at panel B). The SW extension does not contribute significantly to the spectral asymmetry, but does to the distortion of the velocity field.

\subsection{\HI\ kinematic modeling \label{sec:hi_modeling}}
The \HI\ kinematics of M\,100 were modeled using the 3D tilted-ring model software, 3D-Based Analysis of Rotating Object via Line Observations \citep[3D-Barolo, ][]{diteodoro15}. This 3D tilted-ring model is appropriate for our study because it is not affected by the beam smearing, one of the most cumbersome problems in low angular resolution \HI\ image cube. For the titled ring modeling, we adopt the systemic velocity of 1565.4~km s$^{-1}$ (\S~\ref{sec:hi}) and leave the center of rings unconstrained (i.e. to be determined by the modeling software) given the observed morphological lopsidedness in the \HI\ distribution.

Figure~\ref{fig:m100_3dbarolo} shows the rotational velocity, position angle, and inclination derived from the 3D-Barolo modeling of our high-resolution uniformly-weighted \HI\ cube. They are plotted with $\pm 1 \sigma$ uncertainties as a function of radius in panels A, B, and C, respectively. The rotation velocity is about 175 \kms\ at $\sim$1.5~kpc (22\arcsec) and is nearly flat at $V_{rot} \sim 235$~\kms\ from 4.5~kpc (66\arcsec) within uncertainties. Our rotational velocity agrees well with the rotational velocities derived from the 2D tilted ring models \citep[][]{knapen93}. The position angle varies within 130$-$160 degrees, and the inclination is about 23$-$35 degrees, which are consistent with those measured from the velocity field \citep{warmels88, haan08}. The sudden change in the position angle by 20~degrees at a galactocentric radius of 15~kpc can be seen in panel B. This change is likely due to the extended feature towards SW. 

\begin{table}
\centering
\small
\caption{Physical Properties of M\,100}
\begin{threeparttable}
\begin{tabular}{cc}
\hline
\hline
Parameter & Value \\
\hline
Alias & NGC\,4321 \\
Right Ascension (J2000) & 12$^{h}$~22$^{m}$~54.83$^{s}$\\
Declination (J2000) & 15\degree~49\arcmin~30.1\arcsec\\
systemic velocity & 1565.4 \kms \\
redshift & 0.00525 \\
luminosity distance & 13.93 Mpc \\
\HI\ mass & $\rm 3.00 (\pm 0.02) \times 10^{9}$~M$_{\odot}$ \\
D$_{\HI}$ at $\Sigma_{\HI}=1.0$~M$_{\odot}$~pc$^{-2}$ & 6.9\arcmin \\
E(B-V) & 0.023 \\
stellar mass & 4.83$\times$10$^{10}$~M$_{\odot}$ $\rm ^{a}$ \\
SFR & 3.23 M$_{\odot}$~yr$^{-1}$ $\rm ^{a}$ \\
\hline
\end{tabular}
\begin{tablenotes}
\small
\item $^{a}$ Stellar mass and SFR are from \citet{hunt19}. 
\end{tablenotes}
\end{threeparttable}
\label{tab:m100_properties}
\end{table}

\subsection{Kinematically Anomalous \HI\ clouds \label{sec:ahc}}

\begin{figure*}[t]
    \centering
    \includegraphics[angle=0, scale=0.7]{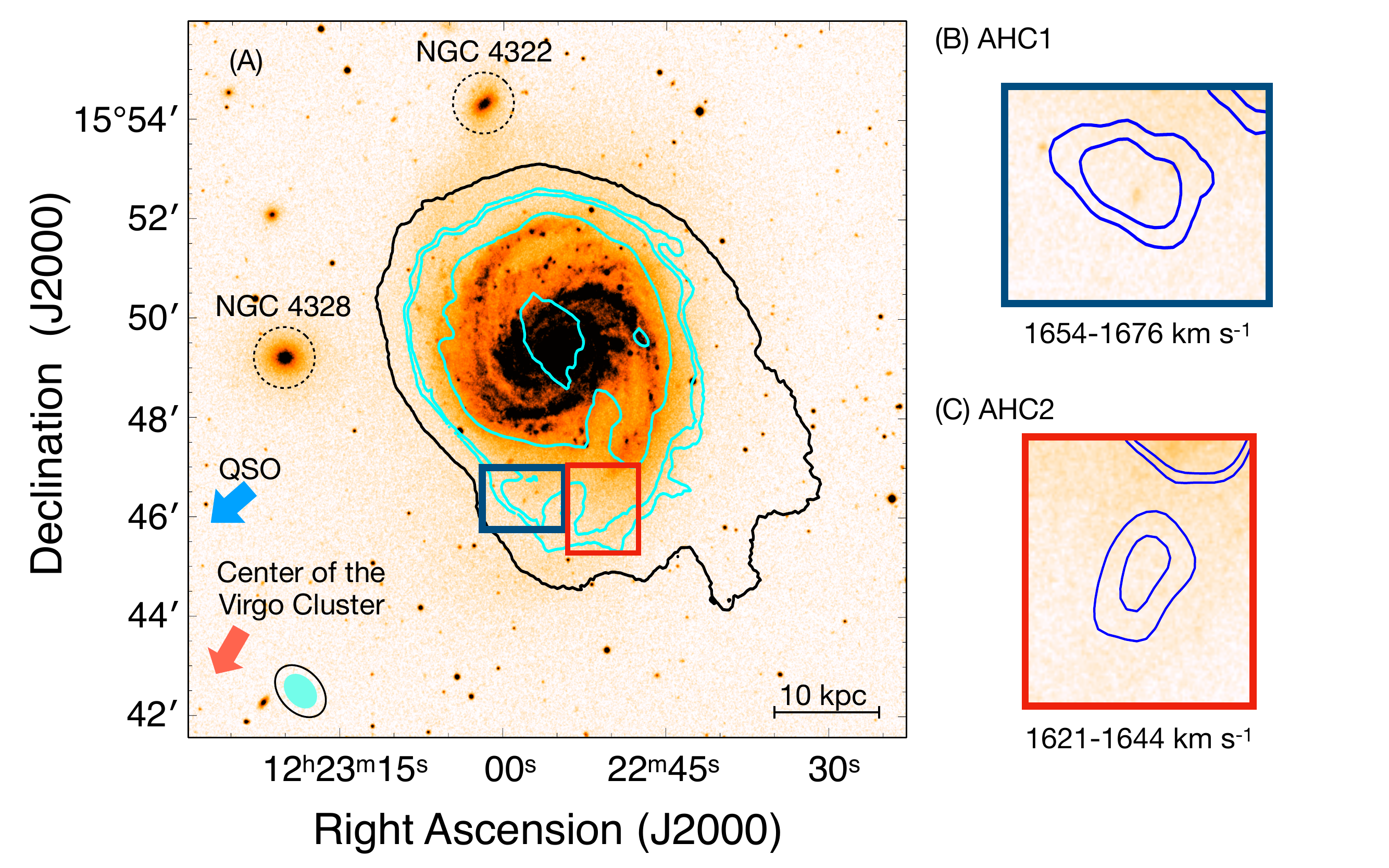}
    \caption{Column density maps of AHCs with the optical image. We present the \HI\ column density maps of M\,100 (panel A) and AHCs (panels B \& C) over the DSS-2 red plate image. In panel A, the black and the outermost cyan contours are the 3$\sigma$ limits in column densities of naturally- and uniformly-weighted maps, which are $1.85 \times 10^{19}$ cm$^{-2}$(black) and $4.56 \times 10^{19}$ cm$^{-2}$ (cyan), respectively. The cyan contours represent 1, 5, 10 $\times 10^{20}$ cm$^{-2}$ in column density of the uniformly-weighted cube. The red arrow points toward the center of the Virgo cluster (12h 30m 49.4s, 12$^{\degree}$ 23\arcmin~ 28.0\arcsec), the blue arrow indicates the location of the QSO SDSS J122330.78+154507.3, and the companion galaxies NGC~4322 and NGC~4328 are marked with the black dashed circle. The synthesized beams are presented in the bottom-left corner, where the large black and cyan filled oval indicate synthesized beam sizes of the naturally-weighted map with 73\arcsec$\times$52\arcsec and the uniformly-weighted map with 44\arcsec$\times$30\arcsec, respectively. Panels B and C present the column density maps of AHCs, measured within the velocity range they were detected. The values are denoted at the bottom of the panels. The boundary color of each panel is the same as that in panel A, i.e., dark blue for AHC~1 and red for AHC~2. Their contour levels are chosen individually for each AHC to highlight their morphologies. The contour levels are $N_{HI}=8.8 \times 10^{19}$, $1.2 \times 10^{20}$~cm$^{-2}$ for AHC~1, and $1.0 \times 10^{20}$, $1.2 \times 10^{20}$~cm$^{-2}$ for AHC~2.}
    \label{fig:opt_hi}
\end{figure*}

We identified two kinematically distinct clouds in M\,100 by visual inspection of the PV diagrams. Figure~\ref{fig:opt_hi} shows the positions of AHCs in the optical image obtained from the Digitized Sky Survey (DSS)-2 red plate, where the AHCs are marked with colored boxes. The colors of the boxes presented in panel A match the boundary colors of panels B and C. The contour levels shown in panel A are the same as those in Figure~\ref{fig:m100_moments}. The contours in panels B and C indicate the total intensity of \HI\ measured by integrating flux densities over the velocity ranges denoted below each panel.

\subsubsection{AHC1 - J122257.27+154614.6}

\begin{figure*}
    \centering
    \includegraphics[scale=0.8]{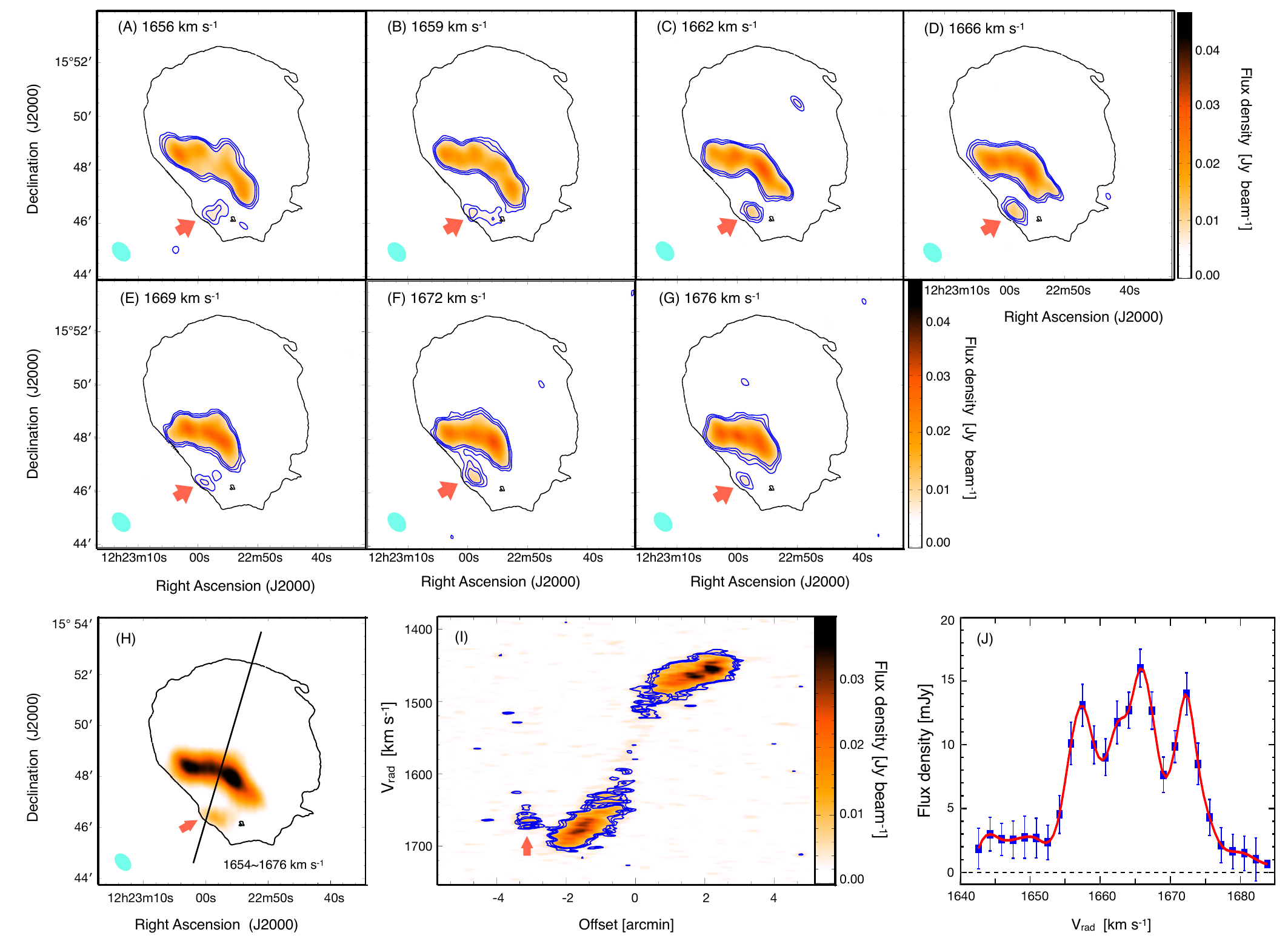}
    \caption{\HI\ properties of the AHC1. We present the channel maps in panel (A)-(G), velocity-integrated flux density  map in panel (H), the PV diagram in panel (I), and the spectrum in panel (J). The contour levels are 3, 4, 5$\sigma$, where $\sigma=1.67$~mJy beam$^{-1}$ channel$^{-1}$ in the channel maps (A-G) and the PV diagram (I). Red arrows in panel (A) through (G) indicate the position of the AHC~1, and the solid line in panel (H) marks the slice for the PV diagram. The spectrum in panel (J) is measured at all channels, but we note that the velocity resolution corresponds to two channels.}
    \label{fig:AHC1}
\end{figure*}

AHC~1 was detected within the radio velocity range of 1654$-$1676 km s$^{-1}$ at R. A. (J2000)=12$^{h}$ 22$^{m}$ 57.27$^{s}$ and Dec (J2000)= 15$^{\circ}$ 46$\arcmin$ 14.6$\arcsec$. This cloud is located at a projected galactocentric radius of 13.4~kpc. 
The channel maps in panels A-G of Figure~\ref{fig:AHC1} show the cloud to be positionally offset from the gas within the disk at those velocities, where the red arrow in each panel indicates the AHC~1's location. The contour levels are 3, 4, and 5$\sigma$ for the $\sigma=1.67$~mJy beam$^{-1}$ per channel. The PV diagram in panel I was generated using the ($\rm kpvslice$) in Karma software package \citep{gooch96} along the line marked in the total intensity map (panel H), where the contours are at the same 3, 4, and 5$\sigma$. 
The PV diagram shows that the AHC~1, which is marked by a red arrow, occupies a distinct location with respect to the rotating disk. We present the \HI\ spectrum in panel J, where the fit (red line) was generated by the spline interpolation in R \citep{rcite}, and the error bars indicates $\pm 1 \sigma$ uncertainties. We note that the spectrum has an effective spectral resolution of 3.3~km s$^{-1}$ corresponding to two channels. Based on the \HI\ spectrum of AHC~1, we estimate the linewidth at 50\% of the peak flux density of $W_{50}=19.1 \pm 4.7$~km s$^{-1}$, the mean velocity of $V_{HI}=1664.6 \pm 4.7$~km s$^{-1}$, an \HI\ mass of ($1.07 \pm 0.15$) $\times 10^{7} \; M_{\odot}$, and a peak column density of $1.63 \times 10^{20}$~cm$^{-2}$. The size of AHC~1 is 87.4$\arcsec \times$54.1$\arcsec$ or 5.9$\times$3.7~kpc. AHC~1 has a velocity deviation of $\Delta V_{sys}=105.3$~km s$^{-1}$ from the systemic velocity. The velocity offset from the disk gas underneath its position based on the interpolation of the edge of the disk in the PV diagram is $\delta v_{pos} \approx$ 40~km s$^{-1}$. \\

\subsubsection{AHC~2 - J122250.66+154547.7}

\begin{figure*}
    \centering
    \includegraphics[scale=0.8]{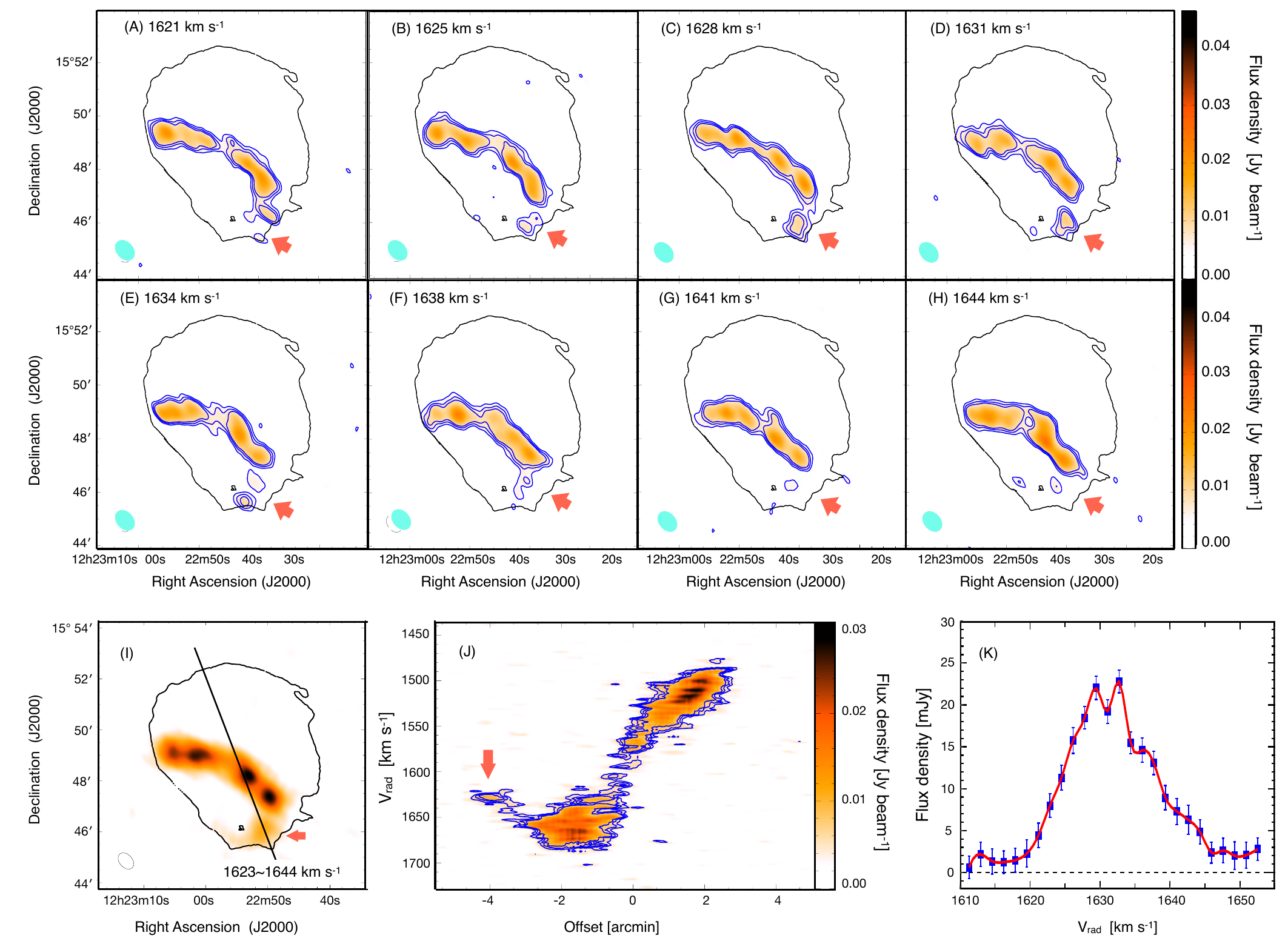}
    \caption{\HI\ properties of the AHC~2. Same as Figure~\ref{fig:AHC1}. }
    \label{fig:AHC2}
\end{figure*}

We identified AHC~2 at R. A. (J2000)=12$^{h}$ 22$^{m}$ 50.66$^{s}$ and Dec (J2000)=15$^{\circ}$ 45$\arcmin$ 47.7$\arcsec$ at a projected distance of 15.5~kpc from the galactic center and radio velocities of $1621-1644$~km s$^{-1}$. The channel maps in panels A-H of Figure~\ref{fig:AHC2} show the cloud separated from the rotating \HI\ disk, even though it might look connected to the rotating \HI\ disk at two channels (panel C and F) due to the low angular resolution of the data. This feature is also prominently distinct from the rotation curve in the PV diagram (panel J) and extends towards lower velocities but at a larger galactocentric distance. 
The spectrum in panel K shows a double-peaked feature, which gives the $W_{50}=14.0 \pm 4.7$~km s$^{-1}$, the \HI\ mass of (1.39$\pm$ 0.14) $\times 10^{7} \; M_{\odot}$, and a peak column density of $1.28 \times 10^{20}$~cm$^{-2}$. The angular size of the cloud is 118.7$\arcsec \times$ 87.8$\arcsec$ corresponding to 8.0$\times$5.9~kpc. The velocity of the AHC~2 is $1631.4 \pm 4.7$~km s$^{-1}$, which yields a velocity deviation of $\Delta V_{sys}=72.1$~km s$^{-1}$ from the systemic velocity of M\,100 and a velocity offset from the rotation curve in the PV diagram (panel J) of $\delta v_{pos} \approx$~40 km s$^{-1}$. \\

We summarize the results mentioned above in Table~\ref{tab:ahc}. The table columns list the ID (A), J2000 coordinates (B), mean velocity (C), \HI\ mass (D), velocity width at 50\% of the peak (E), velocity offset from the rotation curve (F), projected distance from the center of M\, 100 (G), and the physical size (H) of each AHC.

\begin{table*}
\centering
\small
\caption{Properties of the AHCs in M\,100 \label{tab:ahc}}
\begin{tabular}{cccccccc}
\tableline
\tableline
(A) & (B) & (C) & (D) & (E) & (F) & (G) & (H) \\
Name & Coordinates & V$_{HI}$ & M$_{HI}$  & W$_{50}$ & $\delta v_{pos}$ & Distance\footnote{projected distance from the center of M\,100} & size  \\
 & &  (\kms) & ($10^{6}$ M$_{\odot}$) & (\kms) & (\kms) & (kpc) & (kpc) \\
\tableline
AHC~1 & J122257.27+154614.6 & 1664.6 $\pm$ 4.7 & 10.70 $\pm$ 1.50 & 19.1 $\pm$ 4.7 & 40 & 13.4 & 5.9$\times$3.7 \\
AHC~2 & J122250.66+154547.7 & 1631.4 $\pm$ 4.7 & 13.88 $\pm$ 1.44 & 14.0 $\pm$ 4.7 & 40 & 15.5 & 8.0$\times$5.9 \\
\tableline
\end{tabular}
\label{tab:summary}
\end{table*}

\subsection{Star formation activity at the positions of AHCs \label{sec:sf}}

\begin{figure*}
    \centering
    \includegraphics[scale=0.65]{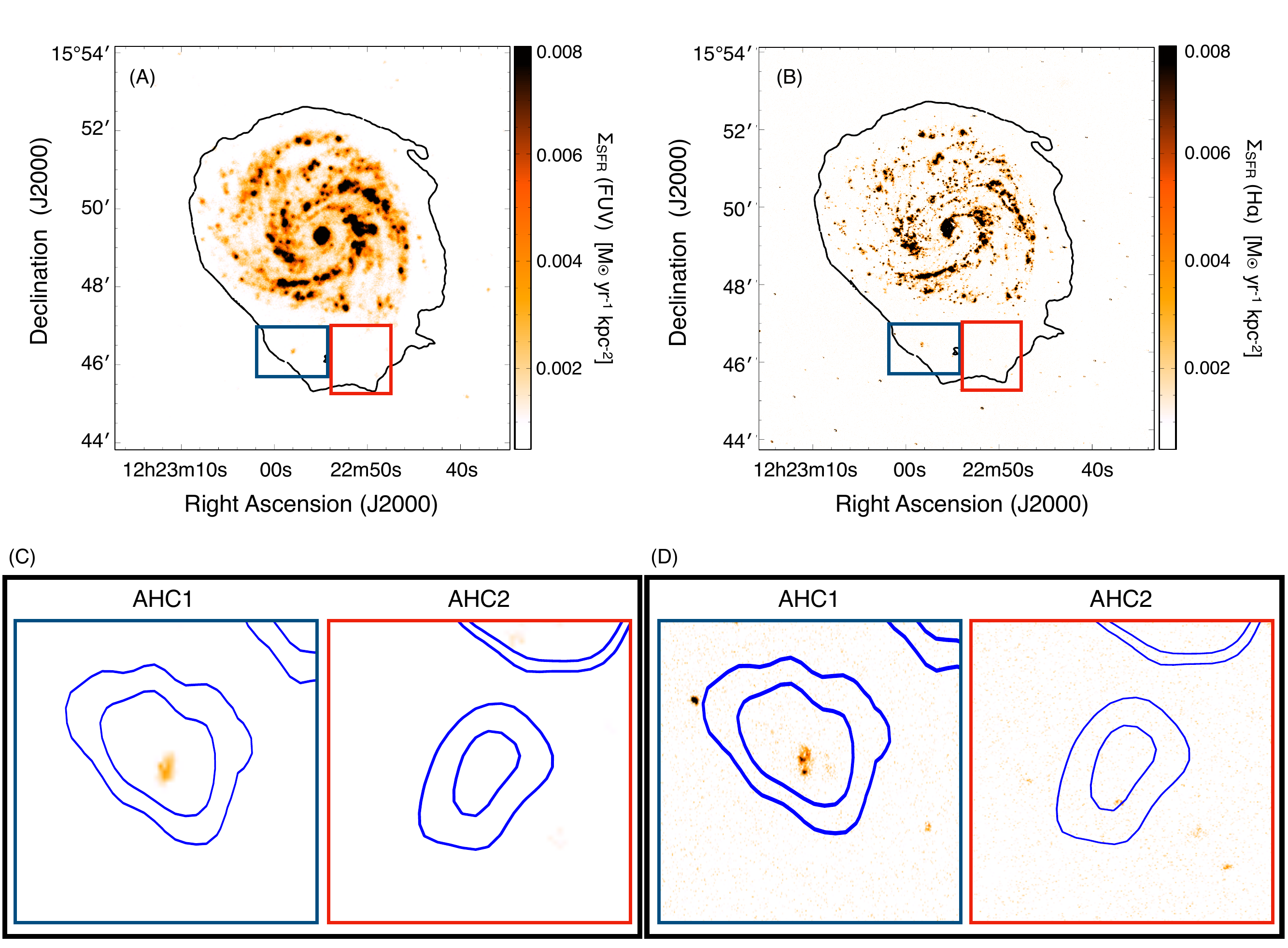}
    \caption{Star formations in the vicinity of the AHCs. The \HI\ contours of the AHCs are plotted over the SFR surface density maps derived from the {\it GALEX} FUV (panel A) and the {\it VATT} H$\alpha$ (panel B) images. The \HI\ contour levels are the same as in Figure~\ref{fig:opt_hi}. The colors of the boxes represent each of AHCs. AHC~1 shows a star-forming region in its vicinity that are capable of generating an outflow, which could explain the origins of the AHCs.}
    \label{fig:ahc_sf}
\end{figure*}

We investigate the stellar contents associated with the AHCs by tracing star formation activity with the {\it GALEX} FUV and {\it VATT} H$\alpha$ images. Figure~\ref{fig:ahc_sf} presents the SFR surface density maps with the \HI\ contour (cyan) showing the \HI\ disk boundary using the uniformly-weighted cube. The boxes inside the maps present the locations of AHCs. Panels C and D show the SFR surface density in each AHC region in FUV and H$\alpha$, respectively. Star-forming regions are detected at the location of AHC~1 in both FUV and H$\alpha$. The presence of H$\alpha$ rules out the possibility of these structures being a background source and help us establish them as star-forming regions at the same redshift as M\,100. Compared to H$\alpha$, FUV is much more sensitive to low-levels of star formation \citep{lee09} and can therefore provide an accurate estimate of the SFRs. By integrating the SFR surface densities above 3$\sigma$ at each band over the star-forming region, we estimate SFRs for AHC~1 as $\rm SFR_{FUV}=$($7.00 \pm 0.17$) $\times 10^{-4}$~M$_{\odot}$~yr$^{-1}$ and $\rm SFR_{H\alpha}=$($5.00 \pm 0.06$) $\times 10^{-4}$~M$_{\odot}$~yr$^{-1}$. These quantities can be translated to the recently formed stellar mass under the assumption of a constant star formation history within the timescale of each tracer.

\subsection{Nature of the Circumgalactic Gas in M100 \label{sec:cgm}}

We do not detect any absorption associated with \Lya\ nor with species tracing higher-ionization states such as \ion{N}{5} and \ion{Si}{4} in the spectrum of the background QSO at an impact parameter of 38.8~kpc corresponding to 1.4~$R_{vir}$. We place the \Lya\ limiting column density of log N(HI) $< 13.24$. This would be highly unusual for a galaxy of the same morphology, luminosity, and SFR as M100 \citep{borthakur16, tumlinson17}. The specific SFR, log (sSFR)=$-10.17$, would put M~100 in the blue cloud \citep{Schiminovich07} and based on the \Lya\ equivalent width to normalized impact parameter ($\rho/R_{vir}$) relationship by \citet{borthakur16}, the expected \Lya\ equivalent width for M~100 is 1.476~$\rm \AA$, whereas the observed limit is 0.094~$\rm \AA$. However, if we consider the \HI\ mass of this galaxy, then the non-detection of cool CGM is consistent with the low gas fraction, $\rm M(HI)/M_{\star}$ of 0.062. It is worth noting that such low gas fractions are not uncommon for galaxies in this stellar mass range \citep{catinella10}. 

In addition, as because M\,100 resides in a cluster, the non-existence of cooler gas in the CGM at an impact parameter of 38.8~kpc is consistent with the findings of \citet{yoon13}. If the circumgalactic gas is in collisional ionization equilibrium with a total hydrogen (\ion{H}{1} $+$ \ion{H}{2}) column density of 10$\rm ^{20}$~particles cm$^{-2}$ along the line-of-sight, and assuming a solar abundance in the CGM, then the non-detection of \Lya\ and \ion{N}{5} would suggest that the CGM is at a temperature greater than 10$\rm ^6$~K. Therefore, it is likely that the cluster medium has impacted even the inner CGM ($\le$~100~kpc from the galaxy), which might eventually lead to cessation of gas condensation from the CGM into the \HI\ disk. Simulations indicate that complete stripping of the CGM is expected for satellites in a cluster like Virgo after a sufficient time \citep{Ayromlou21}. Here, M\,100 is most likely going through a phase of gas starvation/exhaustion  \citep{Donnari21} as it is most likely not able to hold on to its cold CGM reservoir. Therefore, a combination of high halo/stellar mass and the environment is at play to render the CGM of the M~100 devoid of cool gas.

\section{Discussion \label{sec:discussion}}

\subsection{Origin of AHC}

In this section, we discuss the plausible scenarios for the formation and survival of AHCs, and their limitations with our data. Prominent scenarios among them are i) supernova-driven outflow ejected out of the disk or returning gas from the previously ejected outflows \citep[galactic fountain model, ][]{shapiro76}, ii) ram-pressure stripped gas clouds \citep{gunn72}, iii) tidal tails or streams \citep{toomre72, giovanelli81}, and iv) infalling satellites with little or no stellar components \citep{davies04}.

One of the strongest contenders for the origin of AHCs in M\,100 is the galactic fountain model. In this model, star formation-driven outflows containing cool gas are ejected out of the disk, which then combine with the CGM, cool down in the halo, and eventually fall back to the disk. Hydrodynamic simulations showed that this model could reproduce clouds consistent with HVCs of the local galaxies \citep{fraternali08}, including the Complex C \citep{fraternali15} and the Smith cloud \citep{fox16, marasco17}. The positional coincidence of the AHC~1 with the star forming region implies that it might be an outflowing cloud driven by star formation feedback or inflowing clouds that are being re-accreted. It is also located near the spiral arm and has a multi-peaked spectrum indicating multi-component gas clouds with different velocities. Assuming the Salpeter initial mass function \citep{salpeter55}, we estimated that there are 60 massive stars ($\rm > 8 M_{\odot}$) currently in the star forming region associated with the AHC~1. These massive stars may be insufficient to accelerate the AHC~1 with the kinetic energy of $\rm 1.7 \times 10^{53}$~ergs. However, the scenario of supernova-driven outflow is still plausible depending on the star formation history. If the star formation in this region has continued for more than 10~Myrs, then it is likely that multiple generations of massive star feedback have resulted in the energetic of the cloud. We also find that AHCs are connected to the disk gas by bridge-like features at 3$\sigma$ level in some of the channel maps shown in Figure~\ref{fig:AHC1} and \ref{fig:AHC2}. These kinematics that connect the systemic velocity of the AHCs to the bulk of the rotating disk at the position of the AHC indicate that these AHCs are likely recently ejected material from the disk. Such kinematics have been typically seen in the IVCs in the Milky Way Galaxy; the IVCs are found to be spatially and kinematically connected to the rotating \HI\ disk \citep{richter17}.

The spatial proximity between two AHCs implies that they might be influenced by external forces, such as ram-pressure stripping or tidal interactions. Ram-pressure stripping might be one of the most obvious mechanisms for the formation of AHCs in dense environments like the Virgo cluster \citep{gunn72}. That ram-pressure stripping is important in the Virgo Cluster was demonstrated by observations of strong \HI\ deficiencies \citep{chamaraux80, giovanelli83, chung09} and stripped morphologies \citep{oosterloo05, vollmer08}. Figure~\ref{fig:opt_hi} shows that M\,100's disk is truncated at NE and extended toward SW, which is a typical disk morphology of a galaxy influenced by ram-pressure stripping \citep{chung09, vollmer12}. On the other hand, it is hard to determine that ram-pressure stripping indeed produced the AHCs due to the lack of information about the orbital motion of M\,100 within the Virgo cluster. If the truncation and extension of the \HI\ disk is a sign of ram-pressure stripping, then M\,100 should be orbiting perpendicular to the radial motion to/from the center of the Virgo cluster (see the red arrow in Figure~\ref{fig:opt_hi}). In addition, a study of ram-pressure stripping of the Virgo Cluster galaxies classified M\,100 as a pre-stripping galaxy based on its \HI\ morphology, \HI\ deficiency, and \HI\ disk size relative to the stellar disk \citep{yoon17}. The earlier study by \citet{knapen93} argued that ram-pressure stripping is not responsible for the SW extension, because it does not explain the inconsistency between the morphological (NE-SW) and the kinematic (NW-SE) major axes of M\,100. 

Another scenario for the creation of AHCs is via tidal interactions with satellite galaxies \citep{giovanelli81}. The most well-known example is the Magellanic Stream \citep[see a review by ][ and references therein]{donghia16}, where its discrete \HI\ clouds are considered as HVCs \citep{mathewson87}. The HVCs of M~101 might be originated via the tidal interaction with the dwarf companion NGC~5477, which perturbs the \HI\ disk, making the individual gas clouds kinematically decoupled from the rotating disk \citep{combes91}. As for M\, 100, \citet{knapen93} suggested that the passage of NGC~4322 and NGC~4328 could produce the lopsidedness toward the SW, as seen in Figure~\ref{fig:opt_hi}. The passages of NGC~4322 or NGC~4328 might provoke the formations of AHCs. However, we do not see any strong signs of interactions in the \HI\ distribution of M\,100 nor any visible sign in the companion galaxies. A caveat is that our observations may have been too shallow to detect signs of tidal streams in \HI\ around M\,100.

Other possibilities for the origin of these AHCs include remnants of inflating satellite galaxies, especially ones with little stellar content, for example, starless \HI-rich galaxies among ultra-diffuse galaxies \citep{cannon15, leisman17}, starless dwarf galaxies like VIRGOHI21 \citep{davies04, minchin07} or SECCO~1 \citep{bellazzini15b, sand15, bellazzini18}. However, our AHCs are less likely to be inflating satellite galaxies since the parent galaxies of our AHCs must experience a significant mass loss due to the tidal interactions during the accretion given the galactocentric projected distance of 13-16~kpc. According to the relationship between gas fraction and galactocentric distance of local group galaxies \citep{grcevich09}, we expect the parent galaxies of AHCs to have total masses in the range of $\rm 3.2\times 10^{9} - 4.4\times 10^{10}$~M$_{\odot}$. This estimated mass range is an order of magnitude larger than the mass of a dwarf galaxy in general, and such a galaxy would be detected in the optical image. Even if our AHCs originated from disrupted dwarf galaxies, the \HI\ gas clouds could not have survived during the journey to such small galactocentric distances. For example, no \HI\ was found in and around the Sagittarius dwarf galaxy entering the Milky Way Galaxy \citep{koribalski94, burton99}, even though it is located at 20~kpc away from the Galactic center. Therefore, our AHCs are not likely to be dwarf galaxies or \HI\ gas clouds originated from disrupted dwarf galaxies. 

There are also other less likely origins of anomalous velocity clouds. Clouds could be perturbed by a combination of spiral arm gravity and Coriolis force. \citet{knapen93} suggested the possibility of the gas clouds being produced due to the density wave streaming motion in the SW extension of M\,100. However, they indicated that the streaming motion introduces a velocity amplitude of $\sim10-20$~km s$^{-1}$, which is less than half of the observed velocity offsets $\delta v_{pos}$ from the bulk of the rotating \HI\ disk gas at their positions. Therefore, we argue that streaming gas motion would not explain the observed properties of our AHCs.

\subsection{Gas flow rate}
Assuming our AHCs are supernova-driven outflows, we can estimate the gas flow rate in M\,100 and compare it with the star formation rate, which allow us to infer the gas accretion and the galaxy growth. 
According to the bathtub model  \citep[][ and references therein]{somerville15} the total gas flow rates, including the gas accretion and outflow rates, should be comparable to the SFR. We quantify the gas flow rates of our AHCs using the relation given by \citet{richter17}, 
\begin{equation}
\centering
{{dM_{gas,halo}} \over {dt}} = {{M_{gas} v_{in}} \over {d}}
\end{equation}
where, $M_{gas}$ is the mass of the cloud, $v_{in}$ is the infall (or outflow) velocity (corresponding to the velocity deviation from the systemic velocity of M\,100), and $d$ is the galactocentric distance. A caveat of this estimate is that the galactocentric distance is projected from the center of M\,100. As shown in Table~\ref{tab:summary}, the gas flow rates are estimated as 0.002 and 0.086 M$_{\odot}$~yr$^{-1}$ for AHCs. Assuming unidirectionality of the flow, i.e., either all AHCs are tracing inflows or outflows, the amount of total gas flows seen in our data is 0.088 M$_{\odot}$~yr$^{-1}$, much less than the SFR of 3.23~M$_{\odot}$~yr$^{-1}$ for M\,100 \citep{hunt19}. This discrepancy between the estimated gas flow rate and SFR was also observed in the Milky Way, where the gas accretion rate derived from the best-fit \HI\ 21~cm observations of the HVCs is 0.08 M$_{\odot}$~yr$^{-1}$ \citep{putman12}, while the SFR of the Milky Way is about $\sim$0.7-2.3 M$_{\odot}$~yr$^{-1}$ \citep[see a review by][and references therein]{richter17}. This inconsistency implies that our \HI\ 21~cm observations can not trace the entire gas flows. For example, the current \HI\ 21cm observations can not trace the accretion of low angular momentum gas \citep{fraternali08}. In addition, the bulk of the inflowing and outflowing gas is expected to be ionized and hence is not detectable in \HI. For example, in the Milky Way, the gas inflow rate measured from the UV observations of the ionized gas is 0.53  M$_{\odot}$~yr$^{-1}$ \citep[][]{fox19}, which is about 6.6 times larger than that of the neutral gas. Nevertheless, these ``tip of the iceberg" observations allow us to trace their origin and connection to star formation, which is critical to understanding galaxy growth.

\subsection{The fate of AHCs and their connection to the CGM}

Since AHC~1 is associated with an active star-forming region, we speculate that it is outflowing material close to the disk. This conclusion is based on three observations. First, we find AHCs to exhibit small velocity offset of $\lesssim$50~\kms. At that rate, the cloud would reach a distance of $\lesssim$12.5~kpc in 250 Myrs --- the expected lifetime of such neutral clouds moving through the tenuous CGM if they can survive \citep[e.g.][]{armillotta17}. Second, in projection, AHC~1 and AHC~2 both lie on the \HI~disk, indicating that they must indeed be physically close to the disk if we assume an opening angle of 45$^{\circ}$. Third, we do not find any cool CGM at 38.8~kpc. This indicates that the CGM of M\,100 is highly ionized, and the outflowing clouds either do not reach that far or get ionized and assimilated into the hot (T$>10^{6}$~K) circumgalactic gas \citep{heitsch09}. Even for field galaxies, cold gas traced by \HI~ 21cm line emission is extremely rare beyond 20-35~kpc \citep{borthakur16b, borthakur14}, whereas \Lya\ tracing warm partially neutral gas is ubiquitous within the virial radius. Recent semi-analytical modeling by \citet{afruni21} has also shown that star formation-driven outflows may not be the primary source of the cool CGM traced by \Lya\ and other low-ionization metal lines. Therefore, we can conclude safely that AHCs are not likely to reach much further from the disk and would either return to the disk or get assimilated into the hot inner CGM.

The non-existence of a cool CGM (log~N(HI)$\le$13.24) at 38.8~kpc, corresponding to 14\% of the virial radius, is highly unusual for a star-forming galaxy such as M\,100 with a $\rm log(sSFR)\approx -10$. While it is unclear what causes the abnormal CGM composition in M\,100, the Virgo cluster environment likely contributed significantly. Interestingly, the non-existence of a cool CGM has also been reported recently by \citet{Kacprzak21} in a higher redshift field galaxy. These authors claim that the loss of the cool CGM is causing that galaxy to turn ``green" and will eventually lead to the quenching of star formation altogether. This may be happening in M100 as well. We caution, however, that the link between an absence of a cool CGM and quenching is still uncertain. For example, cool circumgalactic gas has been detected in the CGM of red passive galaxies \citep{thom12}.

\subsection{Beam-smearing effect}

Low angular resolution observations like ours may experience spectral linewidth broadening called the ``beam-smearing effect'' \citep{caldu-primo13}, especially in the central regions where rotation curves are steep. In the inner region, \HI\ gas clouds with different velocities are smeared over the beam size, which introduces the spectral smearing. This effect is less significant at the edges of the rotating \HI\ disk, which might explain why we only detected AHCs in the outer disk. However, the AHCs we detected are both on the southern side of the M\,100, which implies that there exist other underlying causes for the origin of these clouds. At present, we have no definitive explanation for the coincidence in the locations of the AHCs, nor whether it is significant. We expect that future VLA observations (approved program 21A-275), with six times higher angular resolution, may find additional AHCs and resolve this issue.

\section{Conclusion \label{sec:conclusion}}
We reported the first detection of two kinematically distinct Anomalous \HI\ Clouds (AHCs) in M\,100 using new VLA data with high spectral resolution. We created multiple image cubes at different spatial resolutions and sensitivities. The \HI\ image cube with a uniform weight function to maximize the angular resolution has a synthesized beam size of 44\arcsec$\times$30\arcsec, an effective velocity resolution of 3.3~km s$^{-1}$ due to hanning smoothing, and an RMS noise of 1.7~mJy beam$^{-1}$ per 1.65 km s$^{-1}$ channel width. We identified the AHCs as kinematically anomalous features that are not co-rotating with the disk. Below, we summarize our findings.

\begin{enumerate}
 \item In terms of the velocity offset, AHCs in M\,100 are analogous to IVCs in the Milky Way. 
 
 \item AHCs have multi-peak spectra and \HI\ masses of $\sim 10^{7} \; M_{\odot}$. They have velocity offsets of about 40~km s$^{-1}$ from the edge of the rotating \HI\ disk, as evident in PV diagrams. AHC~1 is associated with a star formation region in M\,100, which implies that it might be gas clouds ejected by supernova feedback, even though the estimated number of massive stars currently present is insufficient. 

 \item The AHCs detected in M\,100 do not have velocities that are high enough to escape the system. The probable fate of these cloud would be to either get assimilated into the warm/hot circumgalactic medium or rain back down to the disk of the galaxy. The latter possibility is consistent with a galactic fountain model.
 
 \item The gas flow rate estimated from our \HI\ data is 0.088~M$_{\odot}$~yr$^{-1}$, much less than the SFR of M\,100, 3.23~M$_{\odot}$~yr$^{-1}$. This discrepancy is expected as our observations would not detect low angular momentum gas, diffuse/ionized gas, and/or the low column density gas clouds that can contribute to a significant fraction of the flow. 
  
 \item  The inner CGM of M\,100, probed by a QSO sightline at a distance of 38.8~kpc from the center of M\,100, shows no absorption associated with cool circumagalactic gas. The non-detection of \Lya\ provides a stringent limit of neutral hydrogen column density, log~N(HI) $<$13.24. In addition, no metal-line absorbers are seen including high ionization species such as \ion{Si}{4} and \ion{N}{5}, despite the expected high density of the CGM and large pathlength at these small impact parameters. 
    
 \item The small velocity offsets of the AHCs from the disk and the non-detection of the cool CGM by the $HST$ $COS$ at 38.8~kpc suggest that these AHCs are close to the disk and would not contribute significantly to the cool partially-neutral component of the CGM of M100. It is likely that the AHCs will either get assimilated \citep[e.g.][]{armillotta17} or will rain back down to the galaxy before they reach beyond 40~kpc from the disk.

 \item We discussed the plausible mechanisms for AHC formation in M\,100, but no single mechanism fully explains the observations. Rather, these two AHCs likely result from the combination of star formation feedback, ram-pressure stripping, and tidal interactions with satellite galaxies.

 \item Our ability to detect other AHCs is hampered by the beam-smearing effect due to our relatively large synthesized beam size. We expect that future, higher spatial and spectral resolution observations will answer whether our detection rate of AHCs is influenced by the beam-smearing effect. We expect that such higher resolution observations may identify more AHCs with higher significance.

\end{enumerate}

\begin{acknowledgments}

We thank the anonymous referee for highly constructive comments improving this paper. This work was carried out as a part of the VLA-DIISC project with observations at the VLA with the project code of 18A-006. This research is also based on observations made with the NASA/ESA Hubble Space Telescope and NASA's {\rm Galaxy Evolution Explorer}, obtained from the MAST data archive at STScI, which is operated by AURA, Inc., under NASA contract NAS 5–26555. The {\rm Hubble Space Telescope} observations are associated with program GO-14071. Based in part also on observations with the VATT: the Alice P. Lennon Telescope and the Thomas J. Bannan Astrophysics Facility. Part of this research was performed in part at the Jet Propulsion Laboratory, California Institute of Technology, under contract with the National Aeronautics and SpaceAdministration. We give many thanks to Jacqueline van Gorkom, Aeree Chung, and Christian Finlator for useful discussions. We also greatly appreciate NRAO personnel for their hospitality during Hansung B. Gim's visit to Socorro, NM. HG and SB are supported by NSF Award Number 2009409 and NASA grant FP00013428 administrated by STScI. SB, MP, RJ, and DT are supported by NASA grant 80NSSC21K0643. TH is supported by NASA grant FP00013428 administered by STScI. HG, SB, MP, and RJ acknowledge the twenty-two Native Nations that have inhabited the land for centuries where ASU campus is currently placed.

\end{acknowledgments}

\facility{Jansky Very Large Array, Vatican Advanced Automate Telescope, Galaxy Evolution Explorer, Hubble Space Telescope}

\software{CASA (McMullin et al. 2007),  R (R Core Team 2013),  IRAF (Tody 1993), Karma (Gooch 1996)}

\end{document}